\documentclass[conference]{IEEEtran}
\usepackage[utf8]{inputenc}
\usepackage{graphicx}
\usepackage{booktabs}  
\usepackage{xcolor}
\usepackage[absolute]{textpos}

\usepackage{hyperref}
\hypersetup{pdfborder={0 0 0},colorlinks=true,urlcolor=blue,linkcolor=blue,citecolor=blue,bookmarks=true}
\hypersetup{pdftitle={Artificial intelligence adoption in the physical sciences, natural sciences, life sciences, social sciences and the arts and humanities: A bibliometric analysis of research publications from 1960-2021}}
\hypersetup{pdfauthor={Stefan Hajkowicz, Conrad Sanderson, Sarvnaz Karimi, Alexandra Bratanova, Claire Naughtin}}

\begin{document}

\title{\LARGE\bf Artificial intelligence adoption in the physical sciences, natural sciences, life sciences, social sciences and the arts and humanities: \\
A bibliometric analysis of research publications from 1960-2021}

\author
  {
  Stefan Hajkowicz, Conrad Sanderson, Sarvnaz Karimi, Alexandra Bratanova, Claire Naughtin\\
  ~\\
  \textit{CSIRO, Australia}
  }

\maketitle

\begin{abstract}
Analysing historical patterns of artificial intelligence (AI) adoption can inform decisions about AI 
capability uplift, but research to date has provided a limited view of AI adoption across various 
fields of research. In this study we examine worldwide adoption of AI technology within 333 fields 
of research during 1960-2021. We do this by using bibliometric analysis with 137 million 
peer-reviewed publications captured in The Lens database. We define AI using a list of 214 phrases 
developed by expert working groups at the Organisation for Economic Cooperation and Development 
(OECD). We found that 3.1 million of the 137 million peer-reviewed research publications during the 
entire period were AI-related, with a surge in AI adoption across practically all research fields 
(physical science, natural science, life science, social science and the arts and humanities) in 
recent years. The diffusion of AI beyond computer science was early, rapid and widespread. In 1960 
14\% of 333 research fields were related to AI (many in computer science), but this increased to 
cover over half of all research fields by 1972, over 80\% by 1986 and over 98\% in current times.
We note AI has experienced boom-bust cycles historically: the AI ``springs'' and ``winters''.
We conclude that the context of the current surge appears different, and that interdisciplinary AI 
application is likely to be sustained.
\end{abstract}

~


\begin{textblock}{13.4}(1.3,14.9)
\hrule
\vspace{1ex}
\noindent
\footnotesize
\textbf{{$^\ast$}~Published in:} Technology in Society, Vol.~74, 2023. DOI:~\href{https://doi.org/10.1016/j.techsoc.2023.102260}{\textcolor{black}{10.1016/j.techsoc.2023.102260}}
\end{textblock}

\section{Introduction}

The field of artificial intelligence (AI) is generally considered to have got its name at the 
Dartmouth University conference in the United States in the summer of 1956~\cite{ref01}. Since that time the 
field has experienced some ups and downs but has, overall, grown robustly as covered by numerous 
historical accounts~\cite{ref03,ref06,ref04,ref05,ref02}. However, there has been an explosion of AI activity 
in recent times.

The past few years have seen a surge of investment, research, education, training and scholarly 
publishing in AI and machine learning~\cite{ref07,ref07b}. Since 2017 over 700 AI policy initiatives have been 
launched by over 60 national governments and sub-national jurisdictions~\cite{ref08,ref09}. Collectively, these 
announcements were estimated to include over US\$62 billion of new spending~\cite{ref10}. During 2020, 
private-sector investment in AI increased by a record 9.3\% reaching US\$40 billion~\cite{ref11}. Five out 
of the seven most influential papers announced by Google Scholar for 2020 were about AI~\cite{ref12}; 
afterwards, papers related to the COVID-19 pandemic dominated. During 2017–2020 the number of 
university courses teaching AI increased by 103\% at the undergraduate level and 42\% at the 
postgraduate level~\cite{ref11}.

Out of all industry and economic sectors, the science and research sector is among the earliest and 
most enthusiastic adopters of AI technology. AI is a general-purpose technology that can improve the 
cost-effectiveness, speed, safety and quality of research in practically all fields of endeavour 
\cite{ref13}. However, AI may be more than just useful; it could be paradigm-shifting. Some researchers~\cite{ref14} 
argue that AI will ``reshape the nature of the discovery process and affect the organisation of 
science''. A recent workshop hosted by the Organisation for Economic Cooperation and Development 
(OECD) in Paris~\cite{ref15} examined the potential for AI to address the ongoing productivity slump in 
science where more research effort is being expended to achieve the same, or lesser, outcomes 
\cite{ref16,ref17}.

Understanding patterns of AI adoption can help researchers anticipate the future potential of this 
general-purpose technology and invest wisely in capability uplift. However, while there has been 
much work to explore and understand AI adoption within select fields of research, there have been 
comparatively few studies examining the diffusion of AI technology across all fields of research. 
Prior work has also taken narrowly scoped definitions of AI (e.g., machine learning) and limited 
time-periods relating to the past few decades only. We seek to contribute by analysing the adoption 
and diffusion of AI technology across all fields of physical sciences, natural sciences, life 
sciences, social sciences as well as the arts and humanities over history from 1960 to 2021. We have 
a broad definition of AI encompassing 214 phrases that capture practically every facet of this vast 
technological capability.

We continue this paper by reviewing prior research relating to the application of bibliometric 
analysis to analyse patterns of AI adoption within and across fields. We explain how our study 
contributes. We next describe our methods, including our main data source, The Lens. The Lens may be 
a comparatively new tool for many researchers alongside well-known databases such as Scopus, Web of 
Science and Google Scholar. We describe what The Lens is, how we used it and how it compares to 
existing databases. We then present our results relating to the development, application and 
diffusion of AI across the fields of research over history. This is followed by a discussion 
exploring the implications of AI for approaches to human knowledge discovery. We explore whether the 
AI boom-bust cycles of the past are likely to return and the issue of productivity uplift. The paper 
concludes by arguing that the future impact of AI on knowledge discovery will land somewhere on a 
spectrum from useful through to paradigm-changing for researchers in most disciplines.

\section{Related research and our contributions}

There have been several studies using bibliometric analysis, and related approaches, to examine the 
adoption and diffusion of AI in various contexts. For example, the Stanford University 2020 AI Index 
used the Scopus database of peer-reviewed literature to find that 3.8\% of all publications were on 
the topic of AI by 2019 with steep growth in recent years~\cite{ref11}. This is up from 0.82\% in the year 
2000. The Stanford University study finds the total number of peer reviewed AI publications 
increased almost 12 times over the 20-year period leading up to 2019.

Another recent prior study~\cite{ref14} examined the diffusion of ``neural networks'' (NN) – a subset of AI in 
our schema – across 6 research fields of technology, physical sciences, life sciences, biomedicine, 
health sciences, social sciences, as well as the arts and humanities. The authors tracked adoption 
trends during 1990–2018. They identified 260,459 documents on NN in total based on 30 search 
phrases. They found a ``burst of research activity'' leading up to 2018 in all research fields. They 
concluded that AI would likely reshape the process of scientific discovery and change the way 
science is organised. They also argued that AI will emerge as a new general method of invention.

An earlier study by Frank et al.~\cite{ref18}
examined the extent to which major research fields were cited within AI research.
Using a bibliometric analysis,
it was found that mathematics and computer science were most commonly cited in modern AI research 
references, with fewer references to philosophy, geography and art. The authors argued that AI 
research needed to bring in more of the social sciences, and the arts and humanities, to ensure that 
it would hold relevance to policy makers and society more broadly. A related study using 
bibliometric analysis with Web of Science data examined AI publishing patterns across countries, 
academic institutes, collaboration networks, research sponsors and scientific disciplines~\cite{ref19}. This 
study found that diverse disciplines contributed to the multi-disciplinary development of AI 
technology.

There have been numerous studies using bibliometric analysis into the impacts of AI within 
specialised disciplines. For example, Palos-Sánchez et al.~\cite{ref20} examined 73 articles in Web of 
Science and Scopus in the field of human resources management. Using the Bibliometrix tool they 
found that AI applied to human resources management was growing constantly and this was likely to 
continue into the future. They also found that AI applications within this field were focused on 
topics relating to recruitment and job-applicant selection. They noted an opportunity to expand AI 
research into other sub-fields within human resources management. Other subject-specific 
bibliometric analyses have examined AI application in engineering contexts~\cite{ref21}, healthcare settings 
\cite{ref22}, supply chains~\cite{ref23}, renewable energy~\cite{ref24} and education~\cite{ref25}. These studies generally 
concluded that AI adoption is growing within the given field of research and that it’s enabling and 
changing processes of knowledge discovery. Most studies also pointed towards the likelihood of 
continued increased AI adoption.

Our analysis supports and extends upon the previous research summarised above. We examine the 
development and application of AI across practically all fields of research. Our focus is upon the 
differences and complementarities between research fields; not within a single field of research. We 
have also introduced new methods and datasets to enable complementary insights in three main ways. 
First, our analysis is from 1960 to 2021 which covers a longer timespan than previous studies
(e.g., 1991 to 2020 by Liu et al.~\cite{ref19}, and 1990 to 2018 by Bianchini et al.~\cite{ref14}), 
including the first two AI springs and winters and the early diffusion of AI outside of computer 
science fields in the 1980s. Second, we use a more comprehensive set of 214 AI search phrases 
derived from multiple expert working groups at the OECD~\cite{ref26}.
This compares to 30 phrases used by Bianchini et al.~\cite{ref14}
which relate to neural networks only (a subfield of AI),
and a search strategy by Liu et al.~\cite{ref19}
involving partial phrases which, by our estimate, 
accounts for under half of our AI phrases. Lastly, we examine diffusion across all fields of 
research with a comprehensive classification system widely used by Scopus~\cite{ref27}, called the All 
Science Journal Classification (ASJC). This captures a much wider range of research disciplines at a 
more granular scale. We are able to apply the ASJC consistently over the entire 62 year period by 
using The Lens database~\cite{ref28}.

\section{Methods and data sources}

As the volume, variety and velocity of research publishing continues to grow, bibliometric analysis 
is becoming an increasingly popular and effective method for understanding patterns and trends in 
various fields of research~\cite{ref29,ref30}. Bibliometric analysis can help by handling large amounts of data 
(e.g., scholarly publications or citations) and provide quantitative insights into the structural 
relationships that exist in fields of interest~\cite{ref31}. It has been used to study knowledge diffusion 
patterns in blockchain technology~\cite{ref32}, biotechnology~\cite{ref33} and digital transformation~\cite{ref34}. We used 
the approach for analysing patterns associated with AI adoption over the past 62 years.
Consistent  with the bibliometric analysis procedure developed by 
Donthu et al.~\cite{ref31}, 
this section outlines the methodology for the bibliometric analysis, including the data sources, 
research field taxonomy, AI search strategy and reporting metrics used in the analysis.
 
\begin{figure}[!b]
  \centering
  \hrule
  \vspace{2ex}
  \includegraphics[width=0.8\columnwidth]{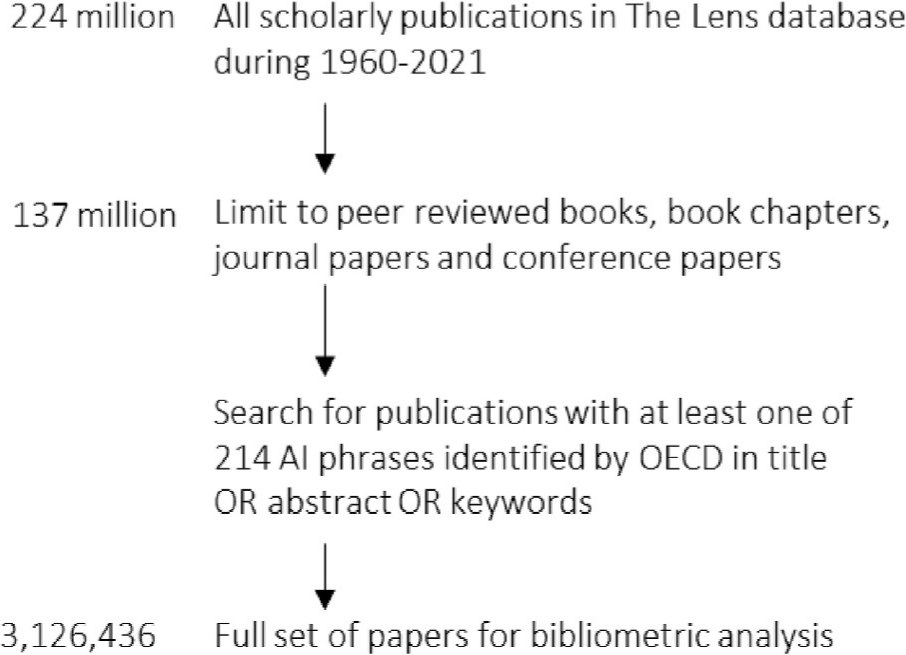}
  \caption{Screening of research publications about artificial intelligence.}
  \label{fig:fig1}
\end{figure}

\subsection{The Lens Database of Scholarly Publications}

Scholarly publication data was sourced from The Lens database (version 8.2), a global database which 
contains over 224 million scholarly publications and over 137 million intellectual property patents 
with records dating back to the 1950s~\cite{ref28}. With early work commencing in 1998, The Lens database 
resulted from a partnership between the Queensland University of Technology and Cambia (both based 
in Australia). Cambia is a not-for-profit organisation aiming to create tools and technologies that 
facilitate knowledge sharing and problem solving. The Lens has a non-commercial nature and receives 
funding from the Bill and Melinda Gates Foundation, the Rockefeller Foundation, and other 
organisations. The Lens database has previously been used for bibliometric analysis for genetic 
science~\cite{ref35} and COVID-19 research~\cite{ref36}.

Scholarly publication data were extracted for records published up until 31 December 2021. Data in 
The Lens was accessed using its graphical user interface (GUI) as well as its application 
programming interface (API) via Python scripts. We used the API to perform customised searches. Data 
in The Lens is sourced from Microsoft Academic Graph~\cite{ref37}, the CrossRef Open Researcher and 
Contributor IDentifier~\cite{ref38}, PubMed~\cite{ref39}, Impactstory~\cite{ref40} and Connecting Repositories (CORE)~\cite{ref41}.

We note numerous repositories containing research publication data which can be used for 
bibliometric analyses (e.g., Scopus, Web of Science and Google Scholar). These databases have been 
reviewed and compared in prior research~\cite{ref42}. The Lens is a relative newcomer and while it is being 
used by researchers for bibliometric analyses~\cite{ref36,ref35} and is well documented~\cite{ref28} it has not yet 
featured in comparative analyses with existing mainstream databases. Nevertheless, we used The Lens 
due to its open-access non-commercial model, comprehensive dataset on scholarly publishing drawing 
upon multiple databases, and high levels of transparency on data provenance. Moreover, some 
databases like Google Scholar do not have a publicly accessible API. This requires manual searches 
which are not feasible in a bibliometric analysis with thousands or millions of publications~\cite{ref43}. 
The Lens provides both an API with comprehensive functionality and a GUI with detailed metadata. The 
Lens removes some of the constraints built into commercial databases which limit transparency and 
data access, which in turn enables improved bibliometric analysis.

\subsection{The All-Science Journal Classification (ASJC)}

We used the Elsevier ASJC taxonomy~\cite{ref27} to examine the diffusion of science. The ASJC has three 
levels. At the most detailed level the ASJC contains 333 unique fields of research. These are 
grouped under 26 subject level fields which are further grouped under 4 fields of physical sciences, life 
sciences, health sciences, social sciences and the arts and humanities. Research publications in The 
Lens are assigned one or more of the 333 third-level fields. The information is derived from the 
International Standard Serial Number (ISSN) descriptions in the Crossref metadata. The pros and cons 
of the ASJC versus other subject matter classification systems have been explored by researchers 
\cite{ref44}. Both classifications are widely used and well-accepted; we consider the ASJC suitable for our 
purposes, but note other classifications are possible.

\subsection{Defining Artificial Intelligence and Identifying Publications for Bibliometric Analysis}

There are several publications reviewing paradigms, approaches and concepts about the definition of 
AI~\cite{ref45,ref46,ref47}. Consistent with previous analyses conducted by the OECD~\cite{ref26}, in this study 
the definition of AI is operationalised via a set of search phrases. These search phrases are used 
to identify publications that are related to AI. We used the list of 214 AI-related phrases provided 
by the OECD~\cite{ref26}. This OECD list of AI-related phrases was developed from a bibliometric analysis of 
publications classified as AI in the Scopus database, which were then interrogated and refined 
further using text mining techniques. The candidate list of AI-related phrases was also validated by 
a panel of AI experts working in business and academic sectors~\cite{ref26}.

Other authors,
such as Liu et al.~\cite{ref19},
have compiled lists of AI search phrases for literature review and bibliometric analysis purposes.
We compared the Liu et al.~\cite{ref19}
and the OECD lists, and found 113 of the 214 OECD phrases had no matching entry
in Liu et al.~\cite{ref19},
76 had a possible matching entry, and 25 had an exact matching entry. These differences are 
due to the OECD list of AI-related phrases including more granular subfields of AI,
such as \textit{pattern recognition} and \textit{computer vision}, which comprise a large share of AI research and publishing.
We adopted the OECD list of phrases as we sought a more comprehensive and inclusive definition of AI.

The initial dataset contained all scholarly publications on The Lens during 1960–2021. We chose 1960 
as the start year as this was relatively soon after the 1956 summer workshop at Dartmouth University 
in New Hampshire where the field of AI was first formally given a name. This dataset was filtered by 
document type, including only records that corresponded to peer-reviewed books, book chapters, 
journal articles and conference papers/proceedings. These scholarly publication records were then 
refined using our search strategy to identify AI-related publications. To be selected as an 
AI-related scholarly publication, a paper needed to contain one or more of the 214 AI phrases 
developed by the OECD in the title, abstract or keywords. A total of 224 million scholarly 
publication records were identified in The Lens database between 1960 and 2021, approximately 87 
million records of which were eliminated as they did not correspond to one of the included document 
types. A further 137 million records titles, abstracts and keywords were screened for AI-related 
terms. The final pool consisted of 3,126,436 records which were included in the bibliometric 
analysis (Fig.~\ref{fig:fig1}).

\section{Results}

\begin{figure}[!tb]
  \centering
  \includegraphics[height=0.65\columnwidth,width=\columnwidth]{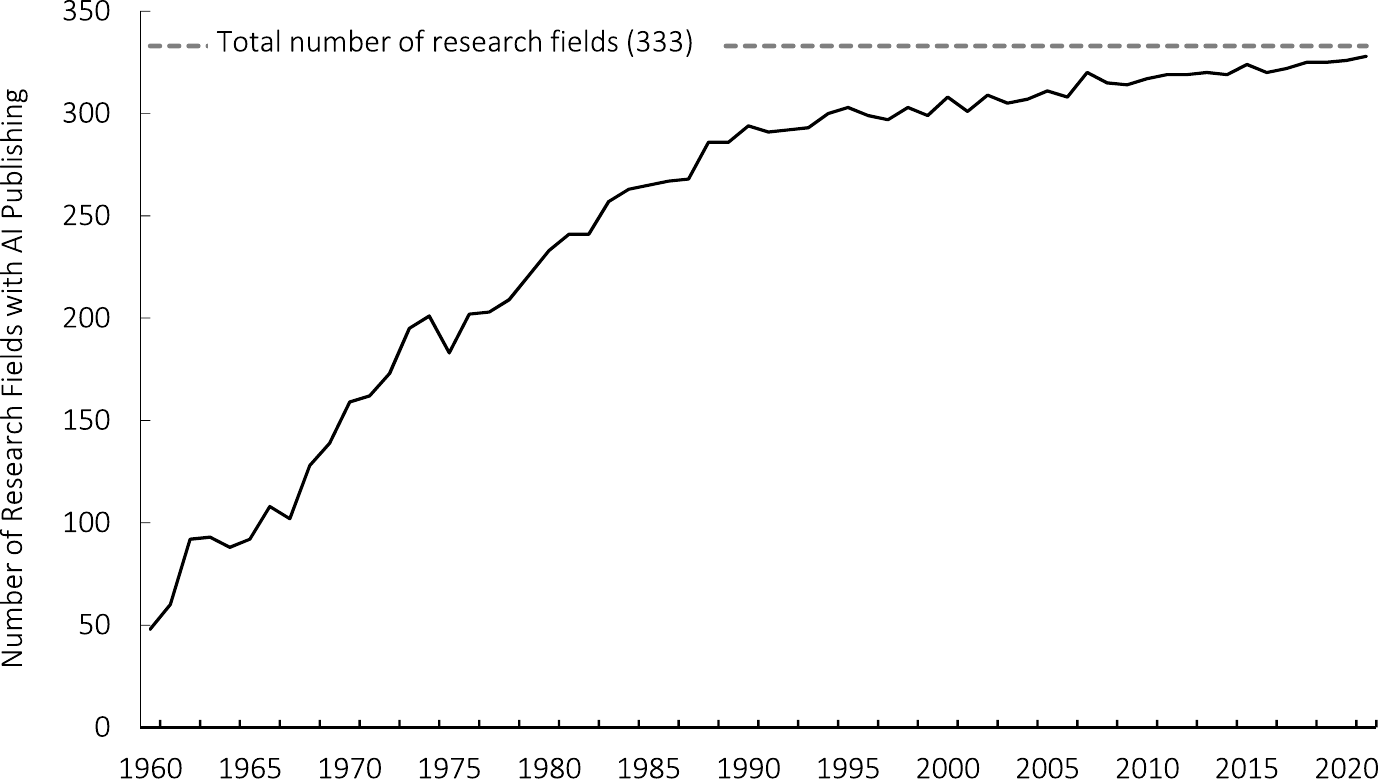}
  \caption{Diffusion of artificial intelligence technology into research fields.}
  \label{fig:fig2}
\end{figure}

In the year 1960 there were 48 research fields with AI-related publications, representing 14\% of 333 
ASJC fields. Most of these were in the fields of computer science, engineering and decision science. 
AI soon spread into other fields; by 1972 over half of all research fields were related to AI. In 
1986 over 80\% of research fields had publications related to AI, and today it is over 98\%.
AI started in the physical sciences and then spread into the life sciences, social sciences, arts and humanities (Fig.~\ref{fig:fig2}).

\begin{figure}[!tb]
  \centering
  \includegraphics[width=\columnwidth]{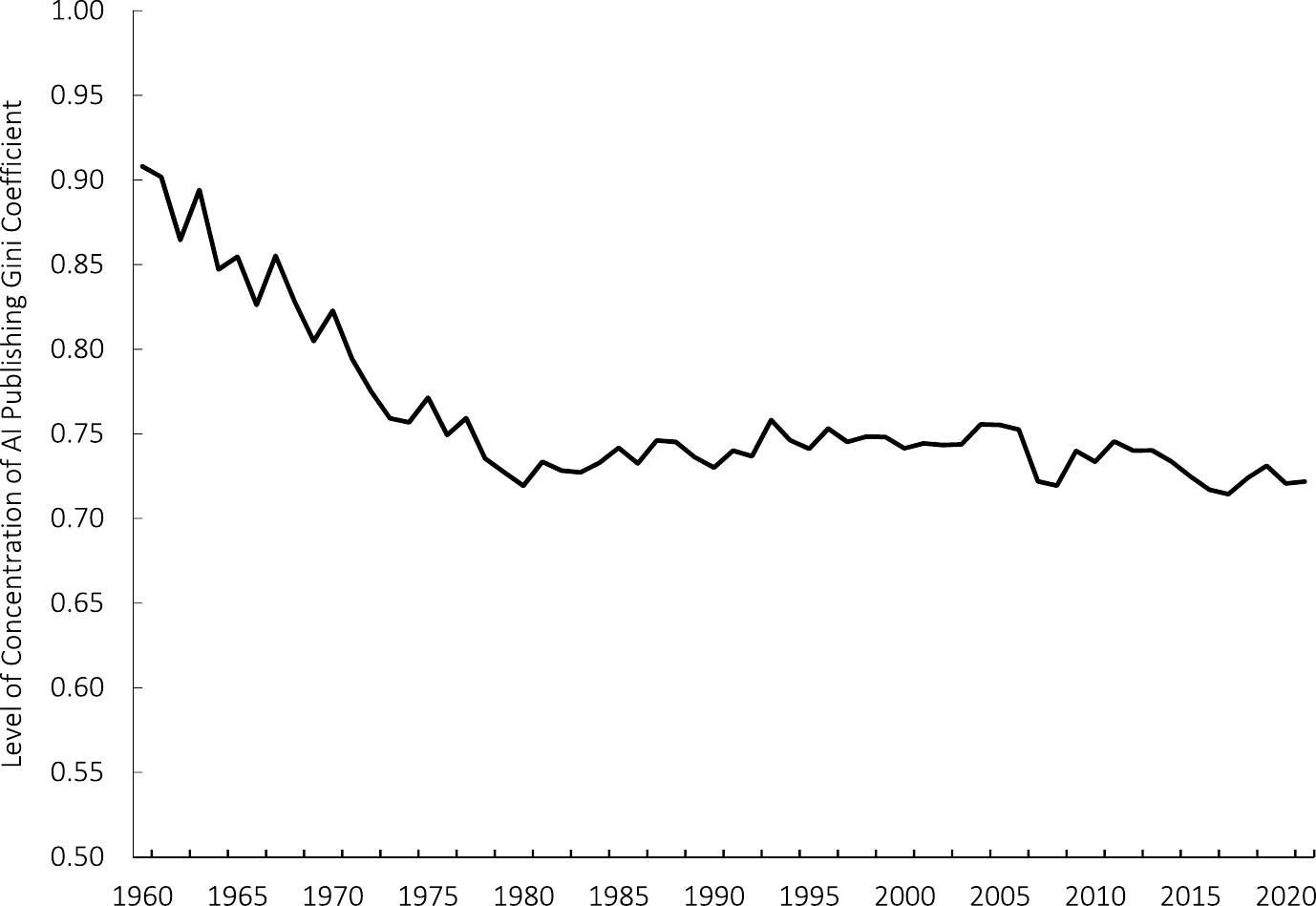}
  \caption{Concentration of artificial intelligence publishing across research fields (Gini coefficient).}
  \label{fig:fig3}
\end{figure}

\begin{figure}[!tb]
  \centering
  \includegraphics[height=0.65\columnwidth,width=\columnwidth]{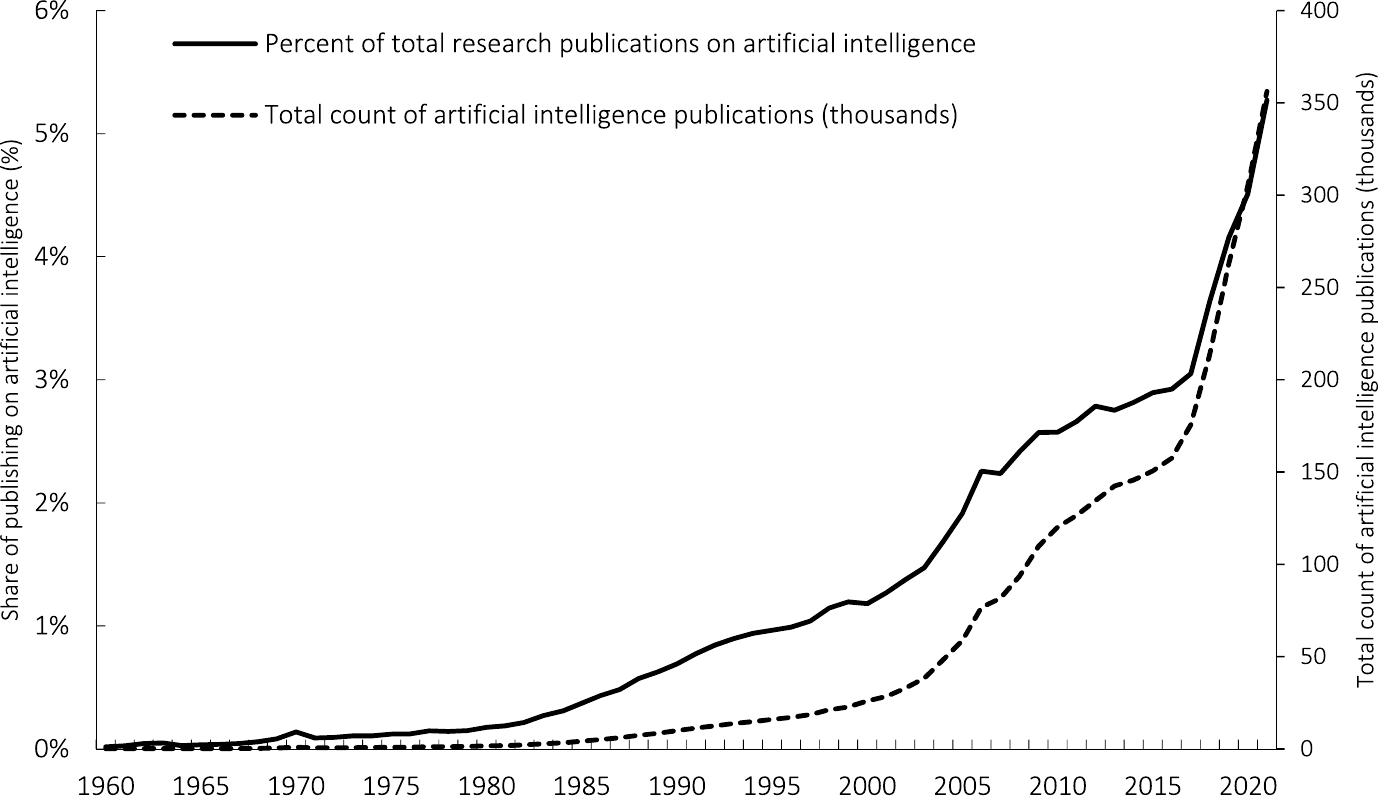}
  \caption{Artificial intelligence publishing intensity and volume over history.}
  \label{fig:fig4}
\end{figure}

AI diffusion can be measured and visualised with the Gini-Coefficient (GC). Often used to measure 
wealth inequality, GC is a statistical measure of how evenly distributed a quantity is among a set 
of categories. The GC value ranges from 0 to 1. When GC~=~0, each category has a perfectly equal 
quantity. When GC~=~1, one category has everything with none in the others. We use GC here to 
determine the spread of AI-related publishing across all 333 fields of research over time. In the 
year 1960, GC~=~0.91 indicating a high concentration of AI publishing in a few fields of research. 
AI did not diffuse from computer science until the 1970s. However, by 1980 the GC had fallen to 0.72 
and has stayed within the range of 0.71–0.76 since that time. One of the reasons it has not fallen 
further is because the computer science field has increased AI publishing intensity and volume at a 
faster pace than any other field. As such, the computer science field has maintained a high 
concentration of AI-related publications in the total publication output (see Fig.~\ref{fig:fig3}).

We define AI publishing intensity as the share (percentage) of total publications that are 
AI-related within a field of research. For all fields of research AI-related publishing started at a 
tiny fraction; 0.02\% of total publication output in 1960. It remained under 1\% until 1995. From then 
until 2017 it increased to 3\%. Over the five-year period 2017–2021 it increased to 5.3\%.
This shows  that most of the AI adoption in research has been happening in the past few years.
More  specifically, over 50\% of the total volume of AI research has been published in the past 5 years. 
The year-on-year growth in AI publishing has averaged at 26\% over the past 5 years, compared to 17\% 
for all preceding history (Fig.~\ref{fig:fig4}).
One of the drivers of recent adoption-growth is the release of accessible AI tools and platforms
which have emerged over the past several years~\cite{ref48b,ref48d,ref48c,ref48},
such as scikit-learn, TensorFlow, Theano, Caffe, Keras, MXNet, mlpack, PyTorch, CNTK, Auto ML, Open NN and H2O.
These and other such tools made AI much more readily available to scientists and researchers
from diverse disciplinary backgrounds.

\begin{figure}[!tb]
  \centering
  \includegraphics[height=0.65\columnwidth,width=\columnwidth]{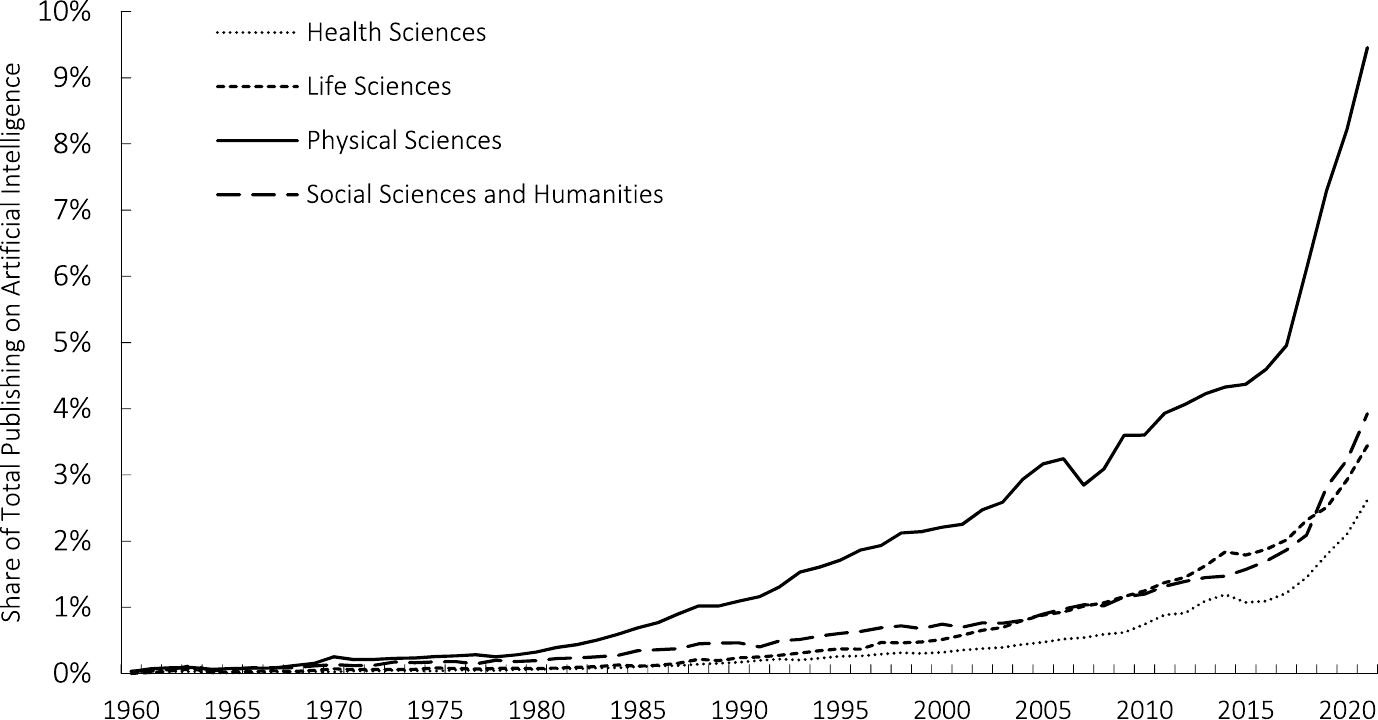}
  \caption{Artificial intelligence publishing in research domains.}
  \label{fig:fig5}
  \vspace{-1ex}
\end{figure}

By examining the science domains and second-level research fields (Fig.~\ref{fig:fig5}, Table~\ref{tab:tab1}),
it can be seen that the physical sciences have, overall, been the largest adopter and developer of AI research. 
However, practically all fields of research show substantial increases in the past 3–5 years.
The five fields of research with the biggest 2020 to 2021 year-on-year increases in AI publishing 
intensity include: dentistry (by 1.9 times); arts and humanities (by 1.4 times);
economics econometrics and finance (by 1.3 times); health professions (by 1.3 times);
and social sciences (by 1.3 times).

\begin{table*}[!tb]
\centering
\normalsize
\caption{\normalsize Artificial intelligence publishing intensity in second-level research fields over history.}
\label{tab:tab1}
\vspace{0.5ex}
\begin{tabular}{l|c|c|c|c|c|c|c|c}
\bottomrule
\textbf{Fields of Research (second-level ASJC)}	& \textbf{1970}	& \textbf{1980}	& \textbf{1990}	& \textbf{2000}	& \textbf{2010}	& \textbf{2015}	& \textbf{2020}	& \textbf{2021} \\ \toprule
Agricultural and Biological Sciences		& 0.1	& 0.1	& 0.3	& ~0.7	& ~1.2	& ~1.5	& ~2.5	& ~2.8 \\
Arts and Humanities				& 0.0	& 0.1	& 0.3	& ~0.4	& ~0.6	& ~0.7	& ~2.3	& ~3.2 \\
Biochemistry Genetics and Molecular Biology	& 0.1	& 0.1	& 0.2	& ~0.4	& ~1.3	& ~1.9	& ~3.2	& ~3.8 \\ \hline
Business Management and Accounting		& 0.5	& 0.5	& 0.9	& ~1.3	& ~2.2	& ~2.6	& ~4.8	& ~5.0 \\
Chemical Engineering				& 0.1	& 0.0	& 0.2	& ~0.7	& ~1.0	& ~1.1	& ~4.1	& ~4.8 \\ 
Chemistry					& 0.0	& 0.1	& 0.2	& ~0.4	& ~0.6	& ~1.0	& ~2.7	& ~3.2 \\ \hline
Computer Science				& 3.7	& 1.9	& 6.9	& 12.4	& 16.0	& 17.1	& 22.7	& 25.7 \\ 
Decision Sciences				& 2.3	& 1.4	& 2.1	& ~4.5	& ~7.1	& ~8.5	& ~9.8	& 11.3 \\
Dentistry					& 0.0	& 0.0	& 0.1	& ~0.3	& ~0.3	& ~0.3	& ~0.9	& ~1.7 \\ \hline
Earth and Planetary Sciences			& 0.1	& 0.2	& 0.5	& ~0.9	& ~1.7	& ~2.5	& ~4.4	& ~5.5 \\
Economics Econometrics and Finance		& 0.0	& 0.1	& 0.3	& ~0.8	& ~0.9	& ~1.1	& ~2.7	& ~3.5 \\ 
Energy						& 0.1	& 0.2	& 0.3	& ~0.8	& ~1.5	& ~2.1	& ~4.5	& ~5.2 \\ \hline
Engineering					& 0.3	& 0.4	& 1.6	& ~3.0	& ~4.4	& ~5.2	& 10.1	& 11.3 \\ 
Environmental Science				& 0.1	& 0.2	& 0.3	& ~0.7	& ~1.3	& ~1.7	& ~2.9	& ~3.3 \\
Health Professions				& 0.1	& 0.2	& 0.4	& ~1.1	& ~1.4	& ~2.2	& ~3.2	& ~4.1 \\ \hline
Immunology and Microbiology			& 0.1	& 0.1	& 0.1	& ~0.3	& ~0.7	& ~1.4	& ~1.9	& ~2.3 \\
Materials Science				& 0.1	& 0.1	& 0.3	& ~0.5	& ~0.7	& ~0.9	& ~4.2	& ~4.1 \\ 
Mathematics					& 0.6	& 0.8	& 1.9	& ~4.9	& ~7.9	& ~9.0	& 12.7	& 14.1 \\ \hline
Medicine					& 0.0	& 0.1	& 0.2	& ~0.3	& ~0.8	& ~1.1	& ~2.2	& ~2.7 \\ 
Neuroscience					& 0.1	& 0.1	& 0.5	& ~1.1	& ~2.3	& ~3.5	& ~5.1	& ~6.1 \\
Nursing						& 0.0	& 0.0	& 0.1	& ~0.2	& ~0.3	& ~0.5	& ~1.1	& ~1.2 \\ \hline
Pharmacology Toxicology and Pharmaceutics	& 0.0	& 0.0	& 0.1	& ~0.3	& ~0.6	& ~0.9	& ~1.7	& ~2.0 \\
Physics and Astronomy				& 0.1	& 0.2	& 0.6	& ~0.8	& ~1.2	& ~1.7	& ~5.6	& ~7.0 \\ 
Psychology					& 0.2	& 0.4	& 0.7	& ~1.2	& ~1.7	& ~2.2	& ~2.7	& ~2.9 \\ \hline
Social Sciences					& 0.1	& 0.1	& 0.3	& ~0.4	& ~0.8	& ~1.2	& ~2.8	& ~3.6 \\ 
Veterinary					& 0.0	& 0.0	& 0.1	& ~0.1	& ~0.2	& ~0.4	& ~1.0	& ~1.1 \\ \bottomrule
\end{tabular}
\end{table*}

The rising intensity of AI publishing in the arts and humanities is a recent phenomenon; throughout 
much of history there has been low, or negligible, AI publishing in these fields. Rising share of AI 
publications in economics is partly being driven by the use of data-driven and AI based approaches 
for econometric modelling and forecasting.
A recent review of machine learning for economic modelling is provided 
in~\cite{ref49}.
The social sciences field captures a highly 
diverse range of disciplines and contains some fields, such as geography, which are early adopters 
of AI and are driving recent increases~\cite{ref50}. Overall, there is an unambiguous pattern in the data; 
AI adoption in research has accelerated in the past few years and AI is now playing an important 
role in most disciplines. It represents a quarter of total research output in computer science.

\vspace{-1ex}
\section{Discussion}

The results from our bibliometric analysis across all fields of research are consistent with 
previous reviews of AI conducted within single, or more narrowly defined, fields of research. We 
found rates of adoption increasing sharply over recent years with the likelihood of continuing into 
the future. This was found in bibliometric analysis of human resources management~\cite{ref20}, engineering 
\cite{ref21}, healthcare~\cite{ref22}, supply chains~\cite{ref23}, renewable energy~\cite{ref24} and education~\cite{ref25}.
Similar patterns of AI adoption rates have been observed 
in~\cite{ref50}, 
which examined the use 
of machine learning in geography and found AI techniques were applied by geographers in cartography, 
spatial statistics and remote sensing during the 1980s. More recent observations have been made in 
literature reviews of AI for dentistry~\cite{ref51}, chemistry~\cite{ref52}, food science~\cite{ref53}, agriculture~\cite{ref54}, 
marine science~\cite{ref55}, econometrics~\cite{ref49} and veterinary science~\cite{ref56}.
There have also been several studies~\cite{ref14,ref18,ref19,ref11} identifying rising multi-disciplinary AI application
albeit with different scope, definitions and data sources to our own.

Our results add to this evidence base, further demonstrating how AI is enabling and most likely 
reshaping longstanding processes of human knowledge discovery. This is demonstrated within numerous 
well-defined and specialised research fields via bibliometric analysis and systematic literature 
reviews. Here we show these AI adoption trends are also observable at higher levels across all 
fields of research in the physical sciences, natural sciences, life sciences, social sciences and 
the arts and humanities. However, there are some critical uncertainties about how AI development and 
application by researchers will unfold over the decades ahead. One critical unknown relates to the 
extent to which the AI boom can be sustained. It is not the first time in history we have seen a 
boom in AI activity which is followed by a bust~\cite{ref13}. The second critical unknown relates to 
productivity gains. Evidence of widespread application is not the same as evidence of productivity 
enhancements. There is every possibility that AI will experience some form of Solow's paradox~\cite{ref57}, 
where information and communications technologies are associated with modest or even negative 
productivity growth due to adaptation challenges.

We first consider the risk of a downturn in AI adoption of opposite, and potentially equal, 
magnitude to the current upturn. Historical analyses suggest that there have been two significant 
surges in AI referred to as ``AI springs'', followed by two downturns called the ``AI winters''. The 
dominant narrative is that the first AI spring was from 1956 to 1974, followed by the first AI 
winter from 1974 to 1981; the second AI spring was from 1981 to 1987 and the second AI winter was 
from 1987 to 1993~\cite{ref02,ref03}. Most historic analyses suggest the causes of the winters were related to 
hype and inflated expectations~\cite{ref06,ref04,ref05}. In the AI springs, there was much confusion and 
misunderstanding about what AI was and what it would be capable of doing. Investment money attracted 
more investment money and AI-related activity increased rapidly. However, when investors realised 
that AI could not deliver on the perceived promises, research funding was substantially and suddenly 
reduced. So, are we currently within a third AI spring likely to be followed by a third AI winter? 
Some researchers have suggested this could be the case, noting that many of the conditions present 
shortly before the two previous winters are happening today~\cite{ref04,ref58,ref59}.

However, we suggest there is reason to believe the next AI winter is not imminent and may not come 
at all, at least not in the same form as earlier winters. First, our analyses show that the current 
surge in AI research far exceeds anything before in history both in the depth (quantity) and breadth 
(diversity) of AI publishing. Second, the current surge in AI applications is coinciding with 
advances in hardware, software and cost-effective cloud computing resources, including the recent 
rise of specialised computing hardware that can handle matrix algebra and support machine learning 
algorithms~\cite{ref60}. Historical technical and financial barriers that previously limited AI adoption and 
use are less significant today. With quantum computing on the horizon, there is the opportunity to 
achieve step changes in what is possible in AI and with AI~\cite{ref62,ref61}.

Next, we can consider the issue of productivity. A key driver for adopting AI in scientific research 
is its potential to boost the productivity of researchers. Productivity declines in science have 
been a focus of scientific conferences~\cite{ref15} and some economists have noted that research outcomes 
have been declining despite increasing research efforts in fields such as agriculture, electrical 
engineering and medicine~\cite{ref16,ref17,ref63}. AI could serve as a general-purpose technology that can be 
applied to lift science and research productivity in all fields of study, similar to the impact of 
electricity in the early 1920s. Here electricity was attributed to productivity gains in the 
manufacturing sector and a period of rapid economic growth across the globe known as the ``roaring 
twenties''~\cite{ref64}. While there is emerging evidence that AI is creating a productivity uplift in 
business~\cite{ref65,ref70,ref66,ref67,ref68,ref69} it is not yet well demonstrated in the science, 
research, innovation and technology sectors. More research is needed to examine this issue.

We also tend to hear more about the AI successes than the failures; it is hard to publish a failed 
AI study. However, the failures do happen and they can be costly. An evaluation of 62 published 
scientific studies using machine learning for COVID-19 diagnosis and prognosis found none of the 
models could be used for clinical purposes as they were subject to methodological flaws or biases 
\cite{ref71}. Other reviews of machine learning applications in COVID-19 diagnostics have similarly 
identified a high prevalence of bias in these AI applications which limits their clinical potential 
\cite{ref72}. AI-based computer vision applications in radiology have also been criticised~\cite{ref73}.
Recognising the shortcomings of AI, and problems for which it is not well-suited,
will be an important part of its adoption in application domains.

Despite the challenges AI capability uplift is likely to be firmly on the agenda for individual 
researchers and research organisations over the coming decade. It is one of the most significant 
disruptors to the scientific method, and processes of knowledge discovery, throughout history. The 
current era is when the adoption rates are steepest. There are two perspectives about how the future 
could unfold. From one perspective AI will be a useful tool which, if used properly, will help 
researchers do what they already do faster, safer, cheaper and better. This is already a huge boon 
to the research profession which is experiencing a productivity slump~\cite{ref16} and urgently needs a boost.
From the other perspective AI is a game changer that reshapes the fundamentals of knowledge discovery.
For example, Gobble~\cite{ref74} explores a roadmap to general artificial intelligence,
and Kitano~\cite{ref75} has proposed the ``Nobel Turing Challenge''
which ``aims to develop a highly autonomous AI system that can perform top-level science,
indistinguishable from the quality of that performed by the best human scientists''.

\section{Conclusion}

We are amid a worldwide surge in AI development and application for research. This is happening in 
the physical sciences, natural sciences, life sciences, social sciences and the arts and humanities. 
While AI has surged in the past, none of the prior events come close to the magnitude and breadth of 
the current situation. While some component of this surge may be hype or trend-following, there is 
likely to be an important and long-lasting substantive component. AI is likely to continue improving 
the speed, cost-efficiency, safety and overall productivity of scientific research. Beyond mere 
efficiency gains, over the coming two decades, AI might fundamentally change the scientific method 
and human approaches to knowledge discovery. The overall implication of this study for researchers, 
and research organisations, is to invest in the many dimensions of AI capability uplift.

~

\bibliographystyle{ieee_mod}
\bibliography{references}

\end{document}